\documentclass[aps,preprint,prd,showpacs,nofootinbib]{revtex4}

\usepackage{latexsym}
\usepackage{amsmath}
\usepackage{graphicx}
\usepackage{subfigure}
\usepackage{dcolumn}
\usepackage{bm}
\usepackage{amssymb}
\usepackage{latexsym}
\usepackage{ulem}
\usepackage{color}
\usepackage[colorlinks,linkcolor=magenta,anchorcolor=cyan,citecolor=blue]{hyperref}

\def\be{\begin{equation}}
\def\ee{\end{equation}}
\def\ba{\begin{eqnarray}}
\def\ea{\end{eqnarray}}

\def\nn{\nonumber}
\def\lf{\left}
\def\rt{\right}

\begin{document}

\title{Chirality oscillation of primordial gravitational waves during inflation}

\author{Yong Cai$^{1}$\footnote{caiyong13@mails.ucas.ac.cn}}
\author{Yu-Tong Wang$^{1}$\footnote{wangyutong12@mails.ucas.ac.cn}}
\author{Yun-Song Piao$^{1,2}$\footnote{yspiao@ucas.ac.cn}}

\affiliation{$^1$ School of Physics, University of Chinese Academy of
Sciences, Beijing 100049, China}

\affiliation{$^2$ Institute of Theoretical Physics, Chinese
Academy of Sciences, P.O. Box 2735, Beijing 100190, China}

\begin{abstract}

We show that if the gravitational Chern-Simons term couples to a
massive scalar field ($m>H$), the primordial gravitational waves
(GWs) will show itself the chirality oscillation, i.e., the
amplitudes of the left- and right-handed GWs modes will convert
into each other and oscillate in their propagations. This
oscillation will eventually develop a permanent difference of the
amplitudes of both modes, which leads to nearly opposite
oscillating shapes in the power spectra of the left- and
right-handed primordial GWs. We discuss its implication to the CMB
B-mode polarization.

\end{abstract}

\maketitle

\section{Introduction}


General relativity (GR) is parity-free. However, in particle
physics and string theory, the anomaly cancellation motivates the
gravitational Chern-Simons (gCS) term
\cite{AlvarezGaume:1983ig}\cite{Jackiw:2003pm} (see
\cite{Alexander:2009tp} for the review),  which is the
leading-order parity-violating correction to GR (see also
\cite{Weinberg:2008hq}\cite{Maldacena:2011nz} for the Weyl
tensor). Thus the chirality of gravity might be a universal
property of the UV-complete theory (see also
\cite{Contaldi:2008yz}\cite{Takahashi:2009wc}).

The primordial GWs \cite{Starobinsky:1979ty}\cite{Rubakov:1982},
predicted by inflation, may induce the B-mode polarization in the
cosmological microwave background (CMB), and carry rich
information about the early universe and the UV-complete gravity
theory, e.g.,
\cite{Cai:2015dta}\cite{Cai:2015ipa}\cite{Baumann:2015xxa}. During
inflation, the primordial GWs generally originate from the
vacuum fluctuation of the metric. The parity-violation of gravity
induces chiral GWs \cite{Lue:1998mq}, i.e., the left- and
right-handed GWs modes propagate with different behaviors.
However, in models of inflation in which a pseudoscalar field
interacts with the gauge field
\cite{Barnaby:2010vf}\cite{Sorbo:2011rz}\cite{Adshead:2013qp}\cite{Adshead:2013nka}\cite{Namba:2015gja}\cite{Peloso:2016gqs}
(see also \cite{Obata:2016xcr}), the gauge field is also
the source of parity-violating GWs. These two
sources seem undistinguishable.

During inflation, the running of the propagating speed of
GWs will result in oscillating shapes
\cite{Cai:2015dta}\cite{Cai:2015ipa}\cite{Cai:2015yza} or a blue
tilt \cite{Cai:2016ldn} in the GWs spectrum. Similarly, the
chirality will also be manifested in the primordial GWs spectrum
\cite{Lue:1998mq}\cite{Alexander:2004wk}\cite{Lyth:2005jf}\cite{Li:2009rt}\cite{Wang:2014abh},
which may bring the TB/EB-mode correlation in CMB
\cite{Saito:2007kt}\cite{Gluscevic:2010vv}\cite{Ferte:2014gja}. Due to the importance
of the parity-violating correction of gravity for the particle
physics and string theory, it is significant to search for its
unique observable fingerprint.

In this paper, we show that if the gCS term couples to a massive
scalar field ($m>H$), the primordial GWs will show itself the
chirality oscillation, i.e., the amplitudes of left- and
right-handed GWs modes will convert into each other and oscillate
in their propagations. This is a novel phenomenon of the chirality
of gravity, which has not been uncovered before.
We illustrate the implications of the chirality oscillation to the
primordial GWs spectrum, and its imprints in the CMB B-mode
polarization.

\section{Chirality oscillation}\label{sec2}

We begin with
the quadratic action for the tensor perturbation \ba
S_{\gamma\gamma}=\int d\tau d^3x\lf[{ M_{p}^2a^2\over8 }\Big(
{\gamma_{ij}'}^2-(\vec{\nabla}\gamma_{ij})^2 \Big)+{\cal
L}_{HD}\rt]\,,\label{Lhh} \ea where $\tau=\int dt/a $ and
$'=d/d\tau$, $M_{p}$ is the Planck scale and ${\cal L}_{HD}$ is
the higher-order derivative  corrections with the parity
violation. Considering only the leading order terms, we could
write \be {\cal L}_{HD}=-{M_p^2\over8}\lf[ {\xi_1\over {\cal M}
}\epsilon^{ijk}(\partial_i\gamma_{jl})'(\gamma_{k}^{l})'
+{\xi_2\over {\cal
M}}\epsilon^{ijk}\partial_i\partial_l\gamma_{jq}\partial^l\gamma_{k}^{q}
\rt]\,,\label{HD} \ee where $\epsilon^{ijk}$ is the Levi-Cevita
symbol, $\xi_1$ and $\xi_2$ are time-dependent parameters,
and ${\cal M}$ is the cutoff scale. The gCS term corresponds to
$\xi_1=-\xi_2$
\cite{Lue:1998mq}\cite{Alexander:2004wk}\cite{Satoh:2008ck}, while
$W\widetilde{W}$ corresponds to $\xi_1=0$
\cite{Weinberg:2008hq}\cite{Maldacena:2011nz}\cite{Baumann:2015xxa},
with $W$ being the Weyl tensor.


In the Fourier space, we have \be
\gamma_{ij}(\tau,\mathbf{x})=\sum_{s=L,R}\int{d^3\mathbf{k}\over
(2\pi)^3}\gamma^{(s)}_{\mathbf{k}}(\tau)p_{ij}^{(s)}(\mathbf{k})e^{i\mathbf{k}\cdot
\mathbf{x}}\,, \ee where the polarization tensors $p_{ij}^{(s)}$
satisfy $p_{ij}^{(R)}{p^{ij}
}^{(R)}=p_{ij}^{(L)}{p^{ij}}^{(L)}=0$,
$p_{ij}^{(R)}{p^{ij}}^{(L)}=2$ and
$ik_l\epsilon^{nlj}{p_{ij}}^{(s)}=k\lambda^{(s)}{p^n_i}^{(s)}$.
The parameter $\lambda^{(L)}=-1$ and $\lambda^{(R)}=1$ correspond
to the left- and the right-handed mode, respectively.

The equation of motion of $\gamma_{\mathbf{k}}^{(s)}$ is \be
{u^{(s)}_{\mathbf{k}}}''+\lf[ \lf(c_{Tk}^{(s)}\rt)^2 k^2-{
{z^{(s)}}''\over z^{(s)} }\rt]u^{(s)}_{\mathbf{k}}=0\,,
\label{eomu1} \ee where $
{u^{(s)}_{\mathbf{k}}}=z^{(s)}{\gamma_{\mathbf{k}}^{(s)}}$ and
$\quad z^{(s)}={a\over2}\sqrt{1-\lambda^{(s)}{k\over
a^2}{\xi_1\over {\cal M}} }$, and the effective sound speed of
different polarization modes
\be c_{Tk}^{(s)}=\lf( 1+\lambda^{(s)}{k\over a^2}{\xi_2\over {\cal
M}} \rt)^{1/2}\lf( 1-\lambda^{(s)}{k\over a^2}{\xi_1\over {\cal
M}} \rt)^{-1/2}. \ee Here, $\xi_1<a$ is required, otherwise the
ghost modes will appear at the cutoff scale $k/a={\cal M}$,
see \cite{Dyda:2012rj}.

Initially, the perturbations are deep inside their horizon, i.e.,
$\lf(c_{Tk}^{(s)}\rt)^2k^2\gg {z^{(s)}}''/{z^{(s)}}$, thus $
{u^{(s)}_{\mathbf{k}}}\simeq
e^{ic_{Tk}^{(s)}k\tau}{/\sqrt{2c_{Tk}^{(s)}k}}$. The power
spectrum of primordial GWs is \be
P_T^{(s)}={k^3\over2\pi^2}\lf|\gamma_{\mathbf{k}}^{(s)}\rt|^2\,,
\quad {aH/ c_{Tk}^{(s)}k}\gg1. \ee The chiral parameter $
\Delta\chi={( P_T^{(L)}-P_T^{(R)}) /( P_T^{(L)}+P_T^{(R)}) }$
reflects the intensity of parity violation of primordial GWs.

When $\xi_1$ or $\xi_2\neq0$, ${\gamma^{(s)}_{\mathbf{k}}}$ with
different polarizations will have different evolutions, the
primordial GWs are chiral. In sting theory, $\xi_1$ and
$\xi_2$ actually are moduli-dependent, see
\cite{Alexander:2009tp}, and the massive moduli fields ($m> H$)
are ubiquitous. The nontrivial evolution (especially the oscillation) of the moduli field indicates the nontrivial variations of $\xi_1$ and $\xi_2$, which may induce oscillation of the chiral term $c_{Tk}^{(s)}$ or ${z^{(s)}}''/z^{(s)}$ in Eq. (\ref{eomu1}). It is this physics that induces the chirality oscillation
of the primordial GWs,
similar to the case in Refs. \cite{Cai:2015dta}\cite{Cai:2015ipa}\cite{Cai:2015yza} where the nontrivial variation (such as oscillation) of the propagating speed of GWs induces oscillations in the GWs modes as well as in the GWs power spectrum.

Generally, the single field slow-roll inflation is thought as the
effective model after all massive modes are integrated out.
However, during inflation, the massive fields could be excited and
oscillate around its minimum.
This will inevitably induce the
chirality oscillation of the primordial GWs, which could be
encoded in the power spectrum of primordial GWs, as will be
showed.

\section{The model }

We will illustrate the idea of the chirality oscillations with a
workable model in detail, in which the background is set as the
inflation with $\epsilon=-\dot{H}/H^2\ll 1$, and $\phi$ is a
massive modulus field with
\be S_{\phi}={M_p^2\over 2}\int d^4x \sqrt{-g} \Big(
-\partial_{\mu}\phi\partial^{\mu}\phi-2V(\phi)+{f(\phi)\over
4M_p^2}R\wedge R \Big), \ee where $R\wedge R=\epsilon^{\alpha\beta\gamma\delta}
R_{\alpha\beta\mu\nu}R_{\gamma\delta}^{~~\mu\nu}$ is the gCS term.
Thus we have $ \xi_1=-\xi_2={ {\cal M}f'\over M_p^2} $ in
Eq. (\ref{HD}), so
 \be z^{(s)}={a\over2}\sqrt{1-\lambda^{(s)}{k\over
a^2 }{ f'\over M_p^2 }  }\,\label{zs1}\ee and $c_{Tk}^{(s)}=1$.
In addition, it is required that ${1\over2}\dot{\phi}^2+V(\phi)\ll
3 H^2, $ and $\lf|-{1\over2}\dot{\phi}^2\rt|\ll
\lf|\dot{H}\rt|$, so that the background is unaffected by
$\phi$.

In this model, if $\phi$ oscillates around the minimum of its potential during inflation, the chiral term ${z^{(s)}}''/z^{(s)}$ in Eq. (\ref{eomu1}) could experience nontrivial oscillation,
which will induce the chirality oscillation of primordial GWs, as explained in Sec. \ref{sec2}.
Here, we would like to point out that the oscillation of $\phi$ around the minimum of $V(\phi)$
is a sufficient condition rather than a necessary one for producing the chirality oscillation of GWs as long as $f$ is $\phi$-dependent, since what plays the key role is ${z^{(s)}}''/z^{(s)}$, which depends also on the shape of $f(\phi)$. For instance, a trivially evolved $\phi$ may also generate the chirality oscillation of GWs if $f(\phi)$ is delicately designed. However, in the following, we will focus on the case where it is the oscillation of $\phi$ that
generates the chirality oscillation of primordial GWs.


\subsection{Numerical Results}

Below, we will numerically show the evolutions of the left- and
right-handed GWs modes and the corresponding spectrum. We define
$\alpha=\ln (a/a_0)$, where there may be a difference of a constant between $\alpha$ and the usually used e-folding number $N$. Then, Eq. (\ref{eomu1}) can be rewritten as \be
u^{(s)}_{\mathbf{k},\alpha\alpha}+\lf(1+{H_{,\alpha}\over
H}\rt)u^{(s)}_{\mathbf{k},\alpha}+{1\over a^2
H^2}\lf(k^2-{{z^{(s)}}''\over z^{(s)} }\rt)u^{(s)}_{\mathbf{k}}=0,
\ee where $z^{(s)}={a\over2}\sqrt{1-\lambda^{(s)}{kH\over
a}{f_{,\alpha}\over M_p^2}}$ and the subscript ``$,\alpha$" is the
derivative with respect to $\alpha$, and the Hubble parameter is
written as $H(\alpha)=H_0 e^{-\epsilon \alpha}$ with $H_0$ set by
the amplitude of curvature perturbation.

Generally, $f(\phi)\sim \phi$, see \cite{Alexander:2009tp} for a
review. To avoid the ghost mode, we require that the
parity-violation only occur in the region around $\phi=0$. Thus we
set 
\be f(\phi)={A_f \phi\over \phi^2+c},  \ee
where $A_f$ and $c$ are
dimensionless, which gives $f_{,\alpha}\approx
g\phi_{,\alpha}$ for $\phi\rightarrow 0$, in which $g={A_f/c}$ is
the coupling coefficient of $\phi$ to the gCS term, and
$f_{,\alpha}\approx 0$ for $|\phi|\gg \sqrt{c}$. During inflation,
the massive modulus field might be excited and then relax towards
its local minimum, and oscillate rapidly around it. Thus for the
simulation, we set \be V(\phi)={m^2\phi^2/2\over 1+(\phi
M_p/{\Lambda})^2}\,,\ee where $\Lambda$ has the mass dimension.
When $|\phi|\gg {\Lambda/M_p}$, \be V(\phi)M_p^2\simeq
{m^2\Lambda^2}\ll \rho_{inf}=3H^2M_p^2 \label{condition}\ee is
constant, which insures that the inflation background is
unaffected by $\phi$. Additionally, the initial condition of $\phi$ is set by
$\phi(\alpha_{ini})=\phi_{ini}$ and $\phi_{,\alpha}(\alpha_{ini})=0$.

Note that, since in our model it is the oscillation of $\phi$ that induces the chirality oscillation of primordial GWs, any potential with a minimum could do the job.
Additionally, the chirality oscillation would present for any $f(\phi)\sim\phi$
when $\phi$ is small. To demonstrate this point, we will also show the oscillation of $\Delta \chi$ with
$f(\phi)=A_f \phi/(\phi+c)$. Although the
shape of the chirality oscillation could be a little model-dependent (which should be attributed to the transition of $f(\phi)$ around $\phi=\pm \sqrt{c}$ or $\pm c$), the phenomenon of chirality oscillation of
primordial GWs is quite general.

\begin{figure}[htbp]
\subfigure[~~$\phi(\alpha)$ and $\phi_{,\alpha}$]
{\includegraphics[width=.48\textwidth]{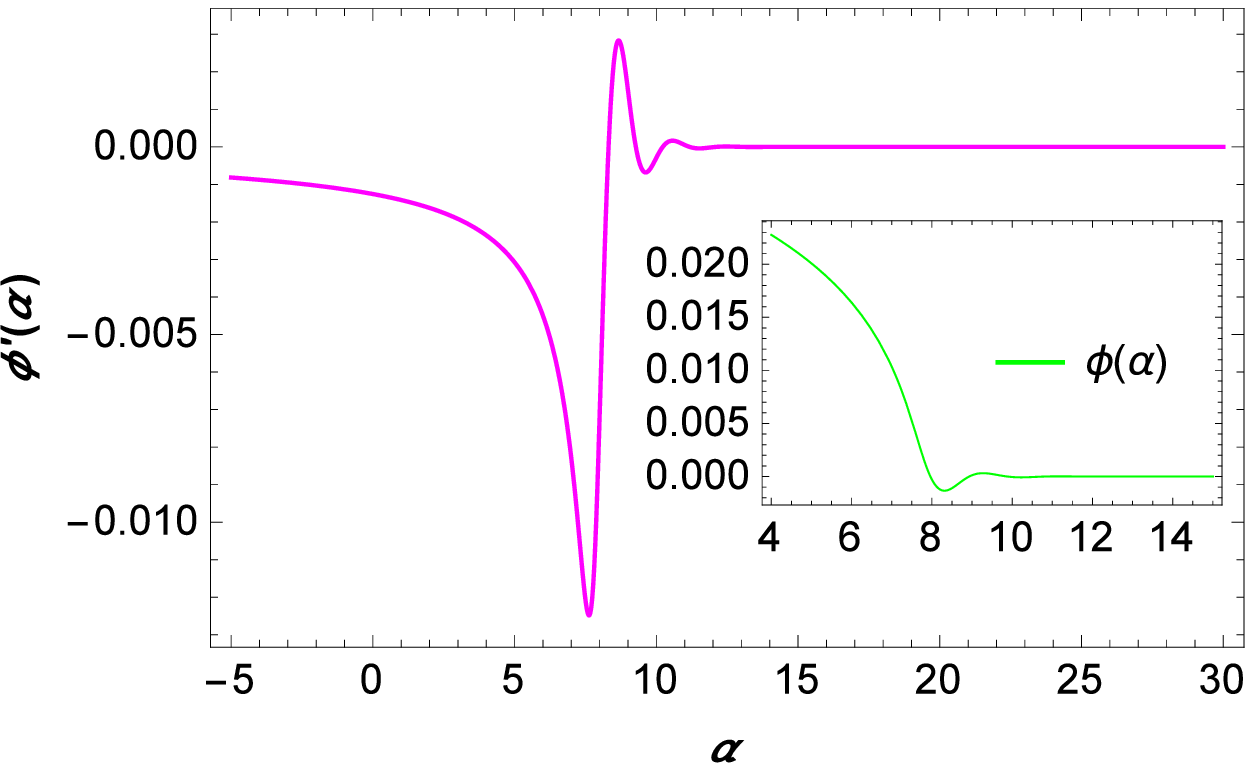} }
\subfigure[~~${ {z^{(s)}}''/z^{(s)} \over a''/a}$ and $\xi_1$ ]
{\includegraphics[width=.47\textwidth]{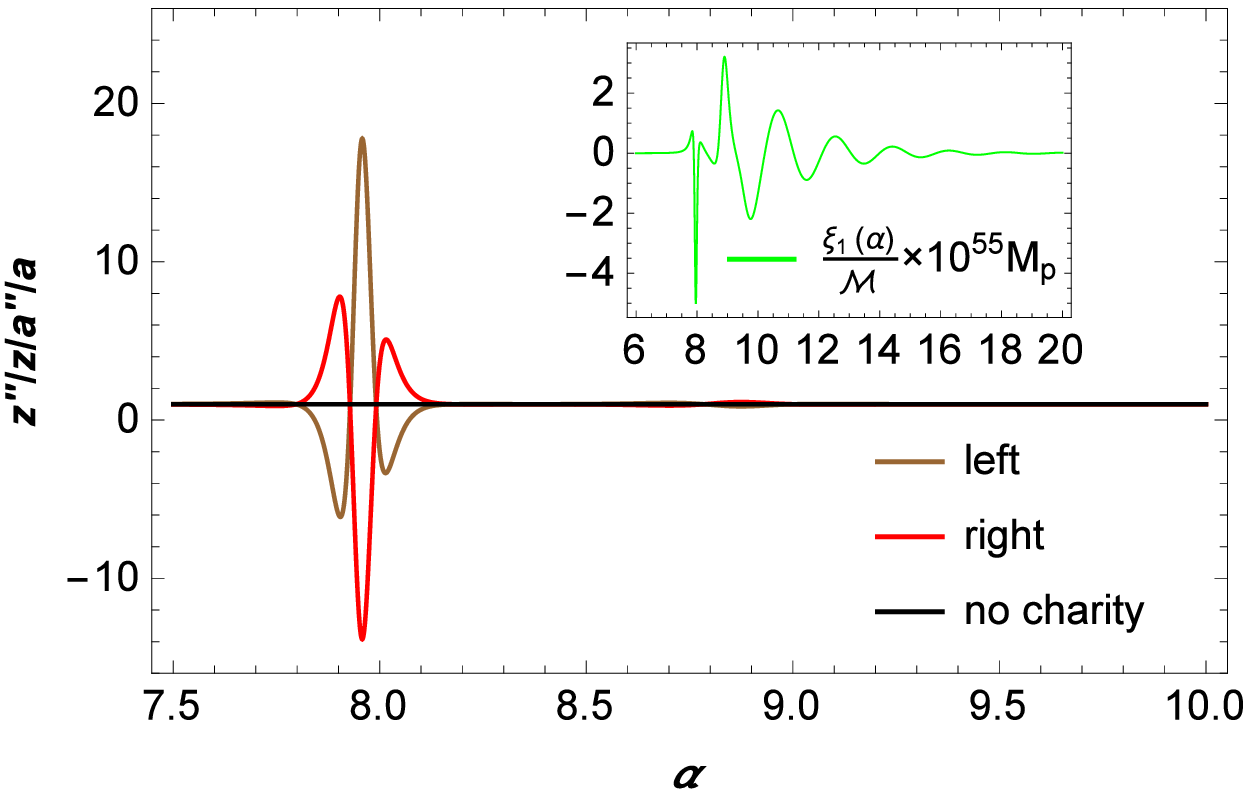} }
\caption{Evolution of $\phi$, $\xi_1$ and ${z^{(s)}}''/z^{(s)}$
when $\alpha_{ini}=-22.8$, $\phi_{ini}=0.044$, $\epsilon=0.003$,  $a_0=10^{-60}$,
$H_0=2.72\times10^{-5}M_p$, $m=9.6\times10^{-5}M_p$,
$\Lambda=0.01M_p$, $A_f=325$,  and $c=4\times10^{-7}$. In (b),
``left", ``right" and ``no chirality" stand for $\lambda^{(s)}=-1$,
$1$ and $0$, respectively.} \label{fig001}
\end{figure}

\begin{figure}[htbp]
\subfigure[~~$|u^{(s)}|$]
{\includegraphics[width=.47\textwidth]{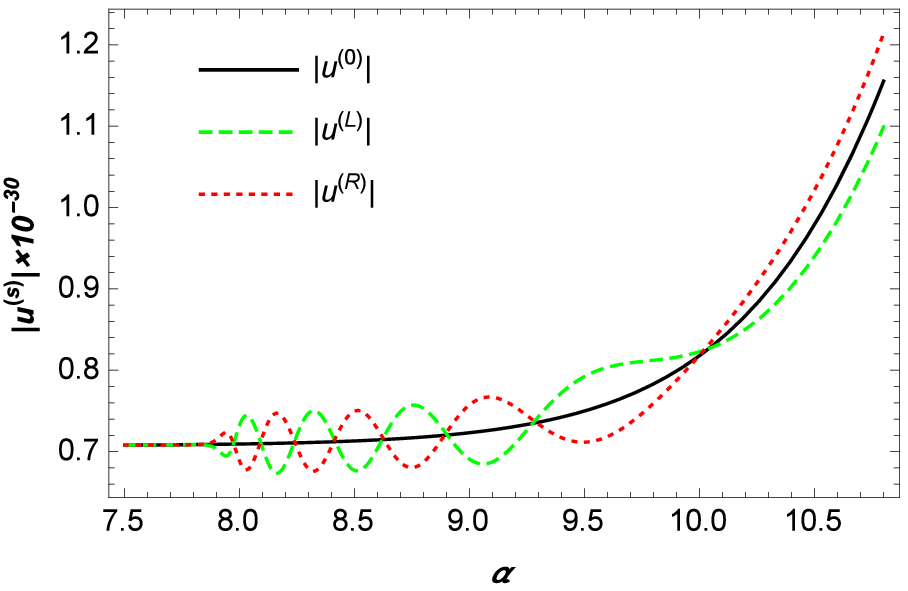} }
\subfigure[~~$|\gamma^{(s)}|/|\gamma^{(0)}|$]
{\includegraphics[width=.47\textwidth]{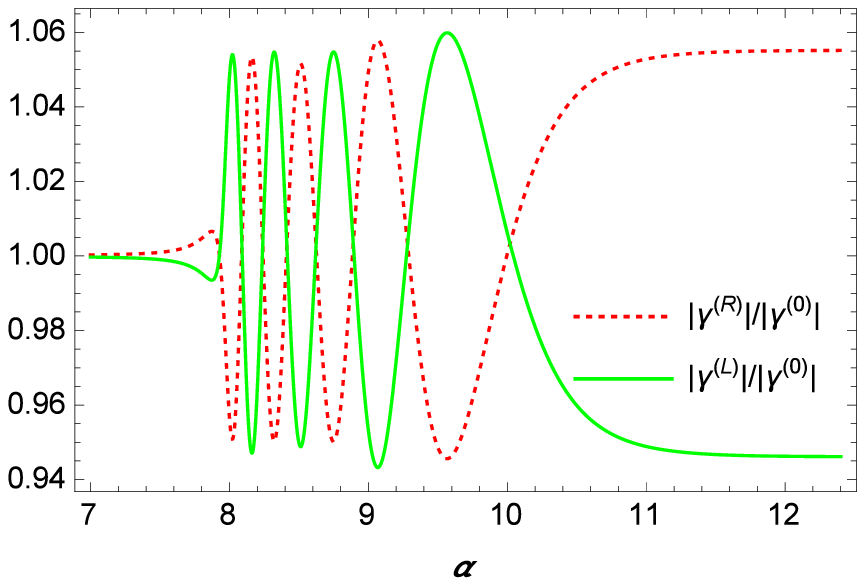} }
\caption{Chirality oscillation of primordial GWs modes. We set
$k=10^{-3}\mathrm{Mpc}^{-1}$, $\alpha_{ini}=-22.8$, $\phi_{ini}=0.044$, $\epsilon=0.003$,  $a_0=10^{-60}$,
$H_0=2.72\times10^{-5}M_p$, $m=9.6\times10^{-5}M_p$,
$\Lambda=0.01M_p$, $A_f=325$,  and $c=4\times10^{-7}$. ``(L)",
``(R)" and ``(0)" stand for $\lambda^{(s)}=-1$, $1$ and $0$,
respectively.} \label{fig002}
\end{figure}

\begin{figure}[htbp]
\subfigure[~~$P_T^{(s)}$  for $f(\phi)={A_f\phi\over \phi^2+c}$ ]
{\includegraphics[width=.42\textwidth]{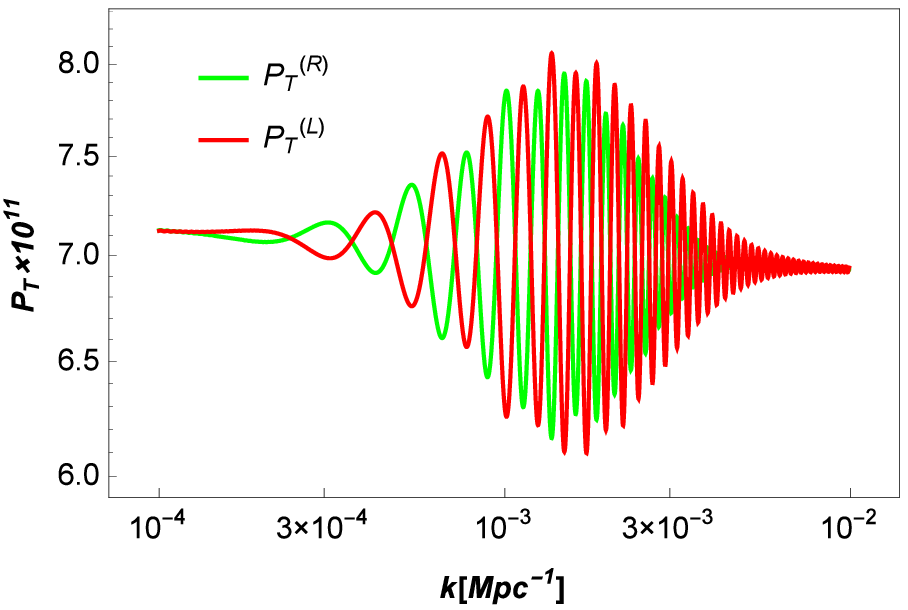} }
\subfigure[~~$P_T^{(L)}+P_T^{(R)}$  for $f(\phi)={A_f\phi\over \phi^2+c}$ ]
{\includegraphics[width=.42\textwidth]{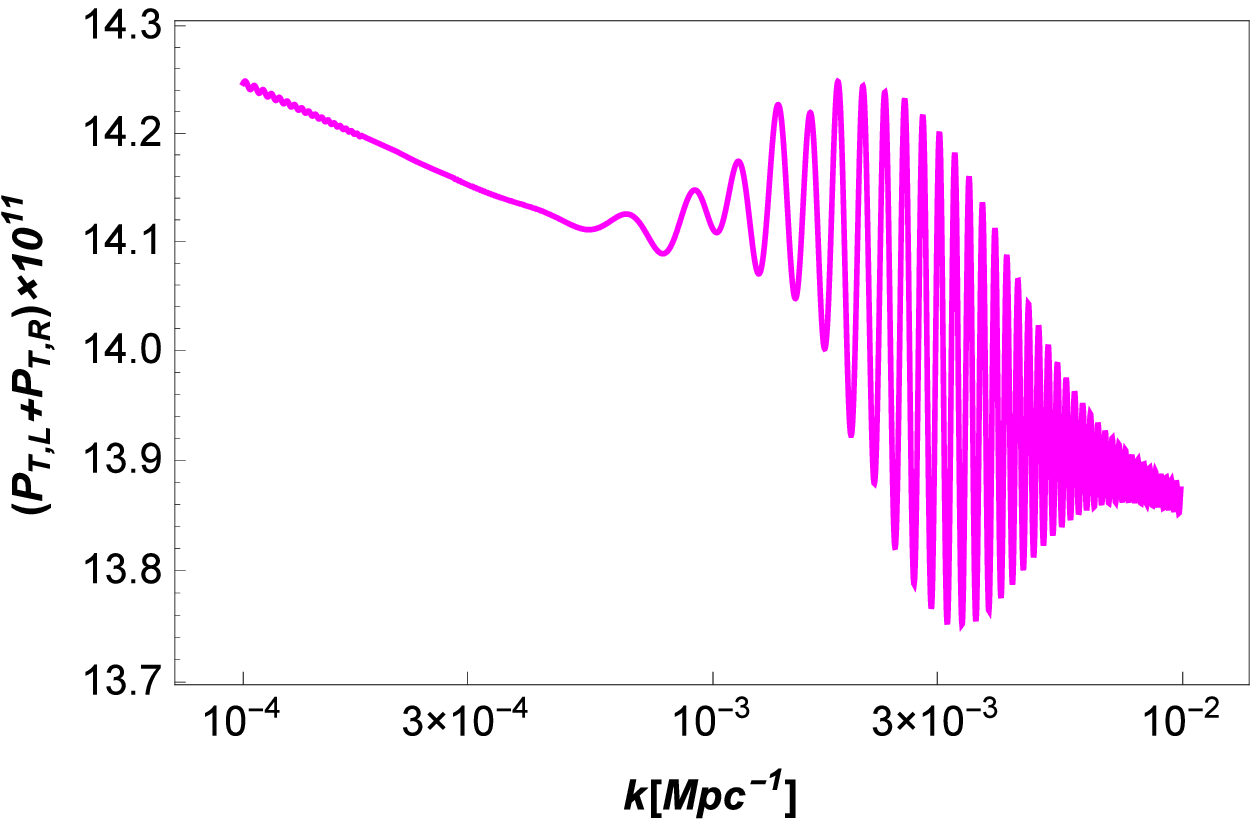} }
\subfigure[~~$\Delta\chi$ for $f(\phi)={A_f\phi\over \phi^2+c}$ ]
{\includegraphics[width=.42\textwidth]{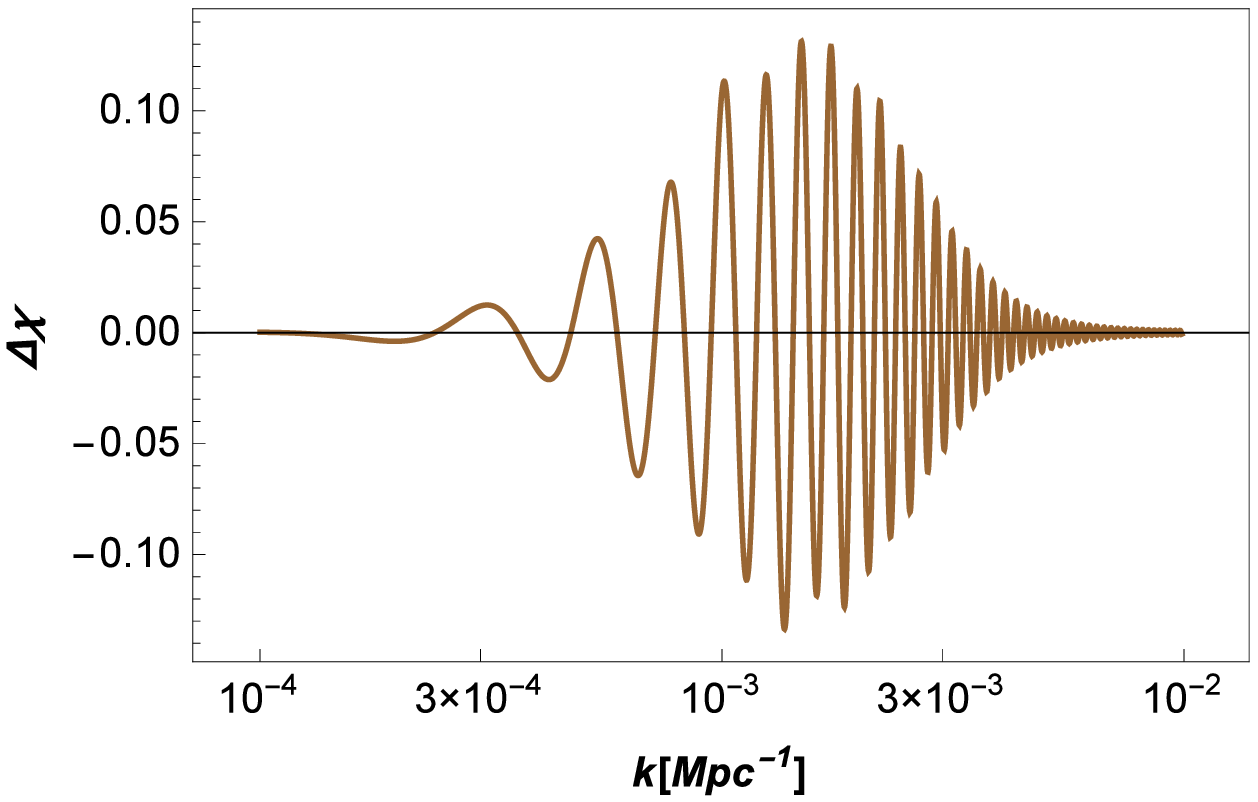} }
\subfigure[~~$\Delta\chi$ for $f(\phi)={A_f\phi\over \phi+c}$]
{\includegraphics[width=.42\textwidth]{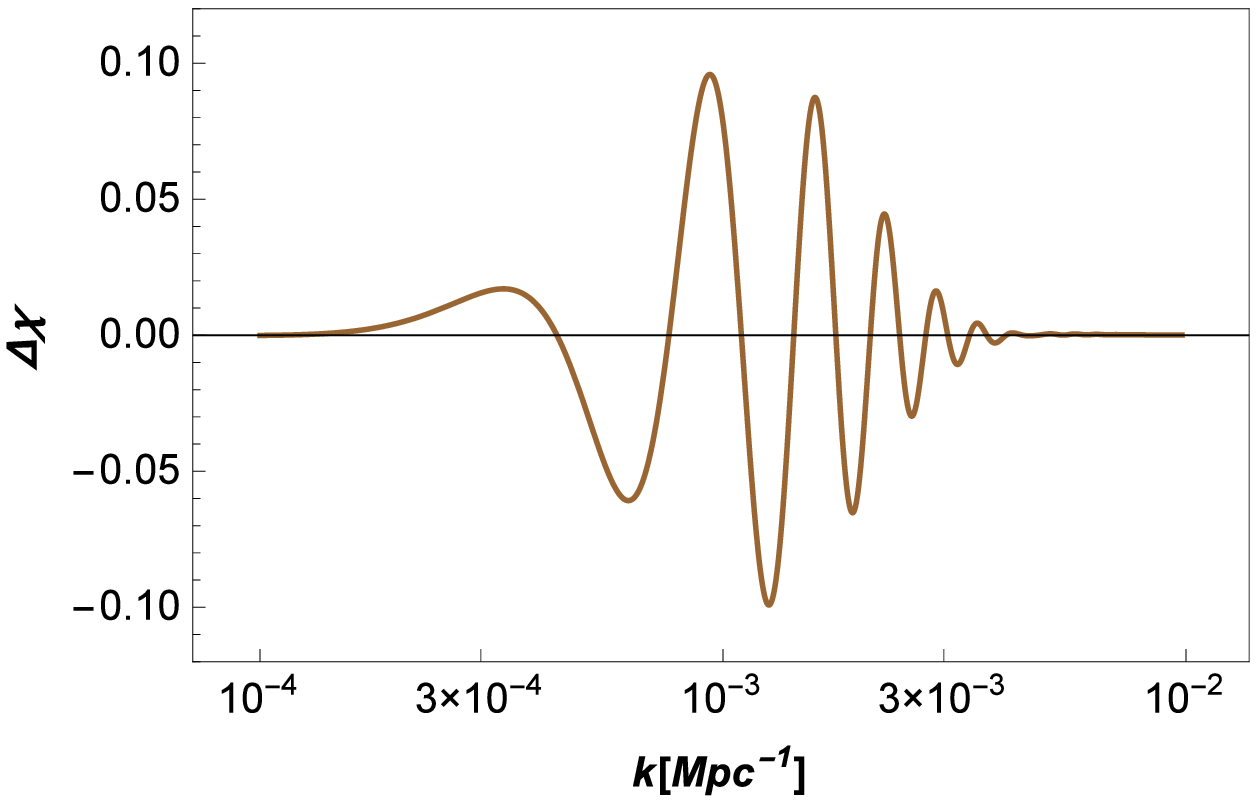} }
\caption{Parity
violating power spectrum of primordial GWs. We set
$\alpha_{ini}=-22.8$, $\phi_{ini}=0.044$,
$\epsilon=0.003$,  $a_0=10^{-60}$,  $H_0=2.72\times10^{-5}M_p$,
$m=9.6\times10^{-5}M_p$ and $\Lambda=0.01M_p$. Note that $A_f=325$ and
$c=4\times10^{-7}$ for (a)(b)(c) while $A_f=1.625\times10^6$ and
$c=2\times10^{-3}$ for (d). ``(L)" and ``(R)" stand for $\lambda^{(s)}=-1$
and $1$, respectively. We have set the cutoff scale at ${\cal
M}\sim10^{-4}M_p$, which corresponds to $k_{cut}\sim a{\cal
M}\sim10^{-2}\mathrm{Mpc}^{-1}$ while $\alpha\approx12$.}
\label{fig003}
\end{figure}

We plot the evolutions of $\phi$, $\xi_1$ (i.e., $f'$) and
${z^{(s)}}''/z^{(s)}$ in Fig. \ref{fig001},
and plot $|u^{(s)}_{\mathbf{k}}|$ and
$|\gamma^{(s)}_{\mathbf{k}}|$ in Fig. \ref{fig002} for the mode
$k=10^{-3}\mathrm{Mpc}^{-1}$, which is at the CMB scale and whose
corresponding frequency is about $10^{-17}\mathrm{Hz}$.
The chirality oscillations of the primordial GWs modes start at around $\alpha=8$, since $\phi$ starts oscillating and ${z^{(s)}}''/z^{(s)}$ starts its nontrivial oscillation at that time.
We see that the chirality oscillation of primordial GWs is encoded in its power spectrum,
as was showed in Fig. \ref{fig003}(a), see also
\cite{Caldwell:2016sut} for the GWs-gauge field oscillation. The
amplitudes of the left- and right-handed GWs mode oscillate almost
(but not exactly) symmetrically, and will eventually arrive at
different values. This difference with respect to the comoving
wave number is reflected in the oscillation of $\Delta\chi$, see
Figs. \ref{fig003}(c) and \ref{fig003}(d). It is also noticed that the broken symmetry
of the evolutions of left- and right-handed GWs modes, i.e.,
$|\gamma^{(L)}|+|\gamma^{(R)}|\neq 2|\gamma^{(0)}|$, actually also
imprint an oscillating fingerprint in $P_T=P_T^{(L)}+P_T^{(R)}$,
as was showed in Fig. \ref{fig003}(b).

\begin{figure}[htbp]
\subfigure[~~$TB $]
{\includegraphics[width=.48\textwidth]{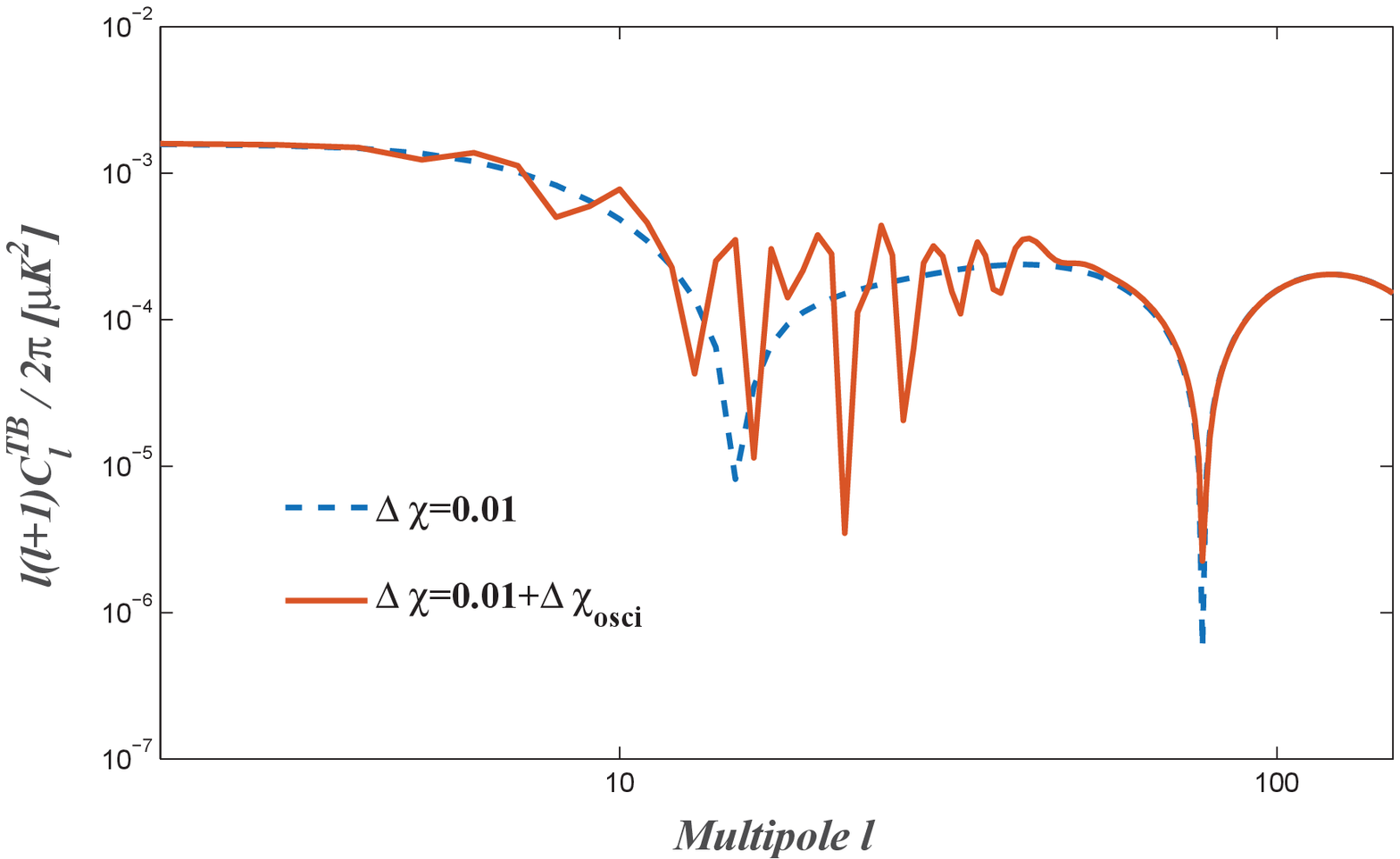} }
\subfigure[~~$EB$]
{\includegraphics[width=.48\textwidth]{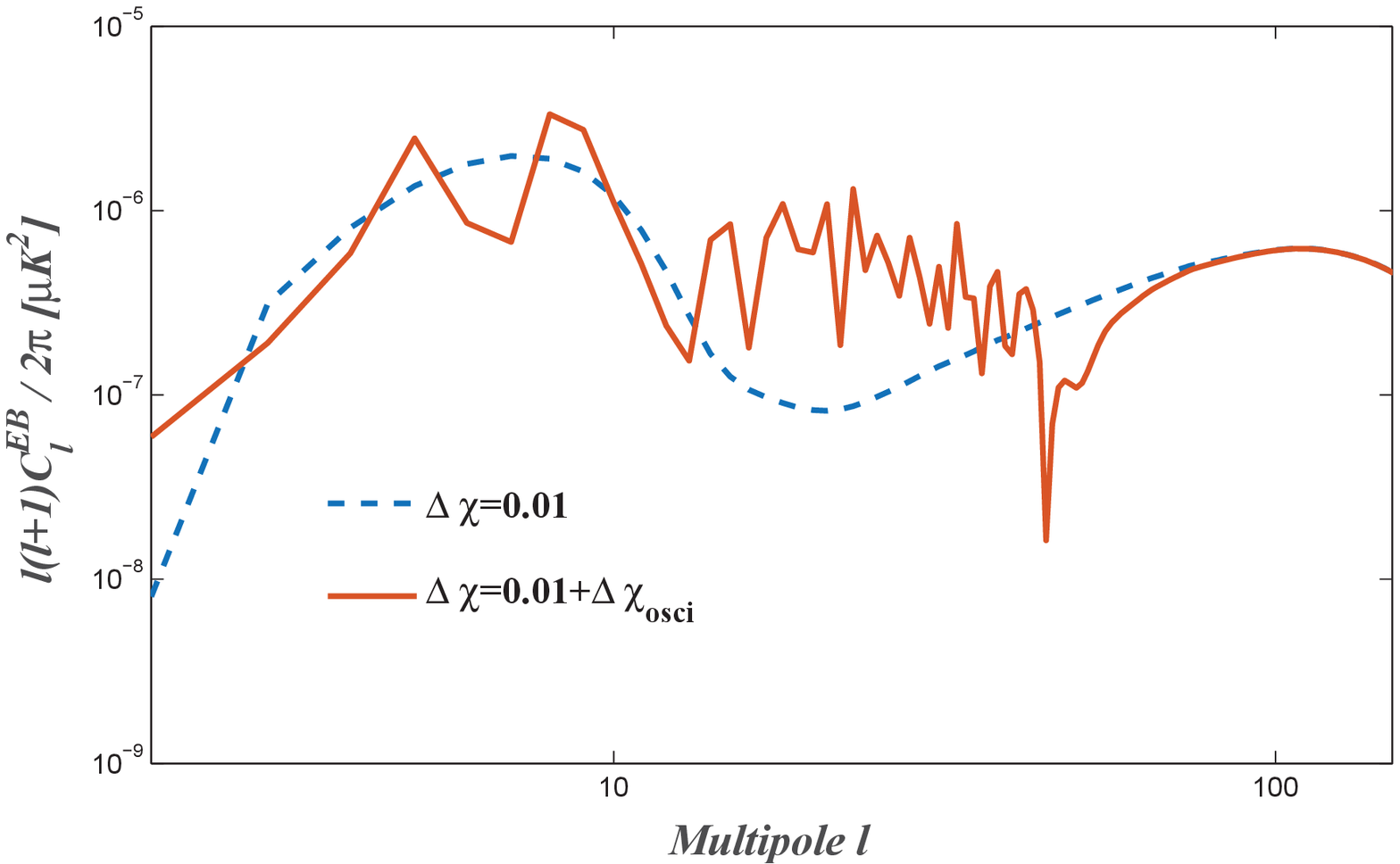} }
\caption{The CMB TB/EB spectrum obtained by modifying the CAMB code. Here, $\Delta \chi=0.01$ is used as a reference to highlight the observational feature of chirality oscillation, and $\Delta\chi_{osci}$ actually equals the $\Delta \chi$ given by
Fig. \ref{fig003}(c). } \label{fig006}
\end{figure}

The chiral GWs will induce non-vanishing TB/EB-mode correlation at
CMB last scatting surface, $C_l^{T/E,B} \sim \int {dk\over k }
\Delta\chi{P}_T [{\Delta_{l}^{ T/E}}(k){\Delta_{l}^{B}(k)}]$
(e.g., \cite{Saito:2007kt},\cite{Gluscevic:2010vv}), if the
modulus field happens to oscillate at the time of $\sim 60$
efolds before the end of inflation\footnote{The perturbation mode with comoving wave number $k\sim 10^{-4}\,\mathrm{Mpc}^{-1}$ exits horizon at about the time of $\sim 60$
efolds before the end of inflation.}.
For parameters used in the numerical calculation, the perturbation mode with $k\sim 10^{-4}\,\mathrm{Mpc}^{-1}$ exits horizon at about $\alpha=8$, where $k\sim aH$ is used for the estimation.
Thus after $a_0$ and $H_0$ are fixed, there is some fine-tuning of the parameters $\phi_{ini}$ and $\alpha_{ini}$, which set the initial condition of $\phi$, so that $\phi$ starts oscillating at about $\alpha=8$.
We plot the TB/EB-mode
spectrum in Fig. \ref{fig006}. We see that the chirality
oscillation of primordial GWs will bring obvious wiggles in the
TB/EB-mode spectrum, which is a novel phenomenon has not been uncovered
before.

\subsection{Analytic Estimation}

With (\ref{zs1}) and the numerical calculation, we approximately have \be {{z^{(s)}}''\over
z^{(s)} }\approx a^2H^2\lf(2-{\lambda^{(s)}kH\over2a
M_p^2}f_{,\alpha\alpha\alpha}\rt). \ee  Then noting
$aH\approx-1/\tau$, Eq. (\ref{eomu1}) can be approximated as
\be {u^{(s)}_{\mathbf{k}} }''+\lf[ k^2-{\lambda^{(s)}A_*k\over2
\tau} -{2\over \tau^2} \rt]{u^{(s)}_{\mathbf{k}} }=0, \ee where
$A_*=H^2 f_{,\alpha\alpha\alpha}/{M_p^2}$ is dimensionless, which
reflects the effect of parity-violating correction of gravity.

Without loss of generality, we assume that before
$\tau=\tau_{mat}$, $A_*=\mathrm{constant}$, and hereafter $A_*=0$, where $\tau_{mat}$ is the comoving time at the matching surface and depends on $\phi_{ini}$ and $\alpha_{ini}$ in the model.
Thus when $\tau<\tau_{mat}$, the solution is \be
{u^{(s)}_{\mathbf{k}1} }=c_{11}\cdot
M\lf({i\over4}\lambda^{(s)}A_*,{3\over2},2ik\tau \rt)+ c_{12}\cdot
W\lf({i\over4}\lambda^{(s)}A_*,{3\over2},2ik\tau \rt),
\label{anau} \ee where $M$ and $W$ are the Whittaker functions.
Initially, the perturbations are deep inside the horizon, i.e.,
$-k\tau\gg1$. Using \be W(x,y,z)\approx
e^{-z/2}z^x\lf(1-{1-4x+4x^2-4y^2\over4z} \rt)+{\cal O}\lf[{1\over
z^2}\rt] \label{wxyz} \ee for $|z|\gg1$ and \be
M(x,y,z)={\Gamma(2y+1)\over\Gamma\lf({1\over2}+x+y\rt)}e^{-i\pi\lf({1\over2}+y-x\rt)
}W(x,y,z)+{\Gamma(2y+1)\over\Gamma\lf({1\over2}+y-x\rt)}e^{ix\pi}W(-x,y,-z),
\ee we have \be c_{11}=0,\qquad c_{12}={ 1 \over \sqrt{2k}
}e^{-{\lambda^{(s)}\over8}A_*\pi }, \ee so that the initial state
satisfies
${u^{(s)}_{\mathbf{k}}}\simeq{1\over\sqrt{2k}}e^{ik\tau}$.

When $\tau>\tau_{mat}$, the solution is \be {u^{(s)}_{\mathbf{k}2}
}={\sqrt{-\pi \tau}\over
2}\lf[c_{21}H^{(1)}_{3/2}(-k\tau)+c_{22}H^{(2)}_{3/2}(-k\tau)
\rt], \ee where $H^{(1)}_{3/2}$ and $H^{(2)}_{3/2}$ are the
$3/2$th order Hankel functions of the first and second kinds,
respectively. Thus we have \be
P_T^{(s)}=P_{T,inf}^{(0)}g(k,\tau_{mat},A_*,\lambda^{(s)}), \ee
where \be g(k,\tau_{mat},A_*,\lambda^{(s)})=|c_{21}-c_{22}|^2,
\label{f}\ee $P_{T,inf}^{(0)}=H^2/\pi^2$ and $\tau_{mat}$
corresponds to the match surface. Then, the chirality parameter \be
\Delta\chi(k,\tau_{mat},A_*)={
g(k,\tau_{mat},A_*,-1)-g(k,\tau_{mat},A_*,1)\over
g(k,\tau_{mat},A_*,-1)+g(k,\tau_{mat},A_*,1)  }. \ee

By requiring the continuities of ${u^{(s)}_{\mathbf{k}} }$ and
${u^{(s)}_{\mathbf{k}} }'$ at $\tau_{mat}$\footnote{When both $a$
and $a'$ are continuous at $\tau_{mat}$, the continuities of $u$
and $u'$ are equivalent to the continuities of $\gamma_\mathbf{k}$
and $\gamma^\prime_\mathbf{k}$.}, with the Tricomi confluent
hypergeometric function $U(a,b,x)$, we obtain
\ba
c_{21}&=&{1\over2}e^{-{\lambda^{(s)}\over8}A_*\pi } \Big[
(4-4ik\tau_{mat})U\Big(1-{i\lambda^{(s)}A_*\over4},4,2ik\tau_{mat}\Big)\nn\\
&\,&+\Big(\lambda^{(s)}A_*(i+k\tau_{mat})-4\Big)
U\Big(2-{i\lambda^{(s)}A_*\over4},4,2ik\tau_{mat}\Big)
\Big]\,,\nn\\
c_{22}&=&{ie^{-{\lambda^{(s)}\over8}A_*\pi }\sqrt{\pi}\over
8\sqrt{-2k\tau_{mat} } } \Big[ -4H^{(1)}_{3/2}(-k\tau_{mat})
W\lf(1+{i\over4}\lambda^{(s)}A_*,{3\over 2},2ik\tau_{mat} \rt)
\nn\\
&\,& +\lf(4k\tau_{mat}
H^{(1)}_{1/2}(-k\tau_{mat})+(4-i\lambda^{(s)}A_*+4ik\tau_{mat})H^{(1)}_{3/2}(-k\tau_{mat})
\rt) W\lf({i\over4}\lambda^{(s)}A_*,{3\over2},2ik\tau_{mat} \rt)
\Big].\nn\\\label{cc}
\ea
%
%

\begin{figure}[htbp]
\includegraphics[scale=2,width=0.55\textwidth]{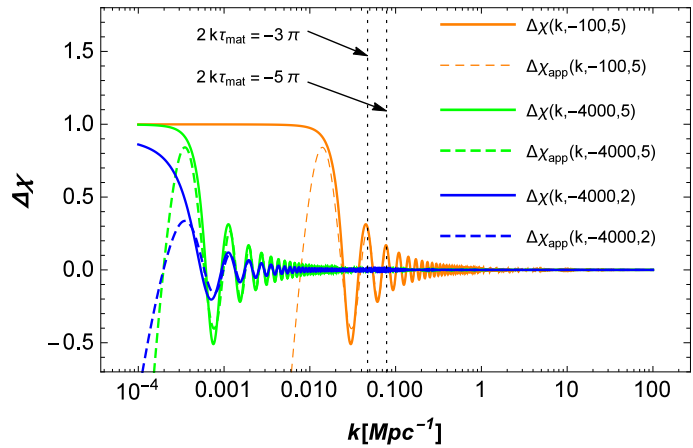}
\caption{$\Delta\chi(k,\tau_{mat},A_*)$ is given by Eqs. (\ref{f}) to (\ref{cc}), and $\Delta\chi_{app}(k,\tau_{mat},A_*)$ is given by Eq. (\ref{chiapp}). The first and the second peak of $\Delta\chi$ correspond to $2k\tau_{mat}=-3\pi$ and $-5\pi$, respectively. } \label{fig05}
\end{figure}

We plot $\Delta\chi(k,\tau_{mat},A_*)$ in Fig. \ref{fig05}. The
amplitude of the oscillation is determined by $A_*$, while the
position of the oscillation is determined by $\tau_{mat}$.
What we are interested in is the oscillation of $\Delta\chi$ when
$k>-1/\tau_{mat}$. Though the amplitude of the oscillation in Fig.
\ref{fig05} is slightly larger than that in Fig. \ref{fig003}(c),
due to the oversimplified assumption we made, we may use the first
or the second peak of $\Delta\chi(k,\tau_{mat},A_*)$ in Fig.
\ref{fig05}, 
as the estimation of the overall oscillating amplitude in Fig.
\ref{fig003}(c).


In the limit $-k\tau_{mat}\gg1$, we have \be
g(k,\tau_{mat},A_*,\lambda^{(s)})=1-{\lambda^{(s)}A_*\over
4k\tau_{mat} }\cos(2k\tau_{mat})+{\cal O}\lf[{1\over
(-k\tau_{mat})^2} \rt], \ee thus
\ba
\Delta\chi_{app}(k,\tau_{mat},A_*) \approx {A_*\over 4k\tau_{mat}
}\cos(2k\tau_{mat}),\label{chiapp} \ea where Eq. (\ref{wxyz})
and $U(x,y,z)= z^{-x}\lf(1-{x+x^2-xy\over z}\rt)+{\cal
O}\lf[{1\over z^2}\rt]$ for $|z|\gg1$ are used. Although
Eq. (\ref{chiapp}) is obtained for $-k\tau_{mat}\gg1$, it is still
able to mimic the oscillation part of $\Delta\chi$ pretty
well\footnote{Up to the next order, \be \Delta\chi=
{4A_*[2k\tau_{mat}\cos(2k\tau_{mat})-3\sin(2k\tau_{mat})]\over
A_*^2+32k^2\tau_{mat}^2-2A_*^2\cos(2k\tau_{mat})}+{\cal
O}\lf({1\over (-k\tau_{mat})^3} \rt) , \ee which doesn't make much
difference for the estimation.}, see Fig. \ref{fig05}.
From Eq. (\ref{chiapp}) we can infer intuitively that the maximum value of the peaks in $\Delta\chi_{app}$ and $\Delta\chi$ corresponds to $\cos(2k\tau_{mat})=-1$. Combined with the numerical results given in Fig. \ref{fig05}, we find the first and the second peak of $\Delta\chi$ correspond to $2k\tau_{mat}=-3\pi$ and $-5\pi$, respectively. 
The peak
values of the first and the second peak of $\Delta\chi$ are
${A_*\over6\pi}$ and ${A_*\over10\pi}$, respectively. When the
values of the parameters we used in Fig. \ref{fig003}(c) are
substituted into Eq. (\ref{chiapp}), we roughly have
$\Delta\chi\sim0.2$, which is consistent with our numerical
result.

Additionally, the parameter $\tau_{mat}$ determines the wavenumber at which the oscillation peak appears, as we can see from Fig. \ref{fig05}. However, since the main purpose of this analytical approximation is to have an estimation of the amplitude of the chirality oscillation, the value of $\tau_{mat}$ is not so important here.

As has been mentioned, $A_*=H^2 f_{,\alpha\alpha\alpha}{
/M_p^2}$, which approximately is $A_*\simeq g{\dddot
\phi}/HM_p^2$. During the rapid oscillating of $\phi$, we
neglect the effect of cosmological expansion, and have $|{\dddot
\phi}|\simeq m^2|{\dot \phi}|\simeq m^3\Lambda/M_p$. Thus \be
A_*\simeq {g m^3\Lambda\over HM_p^3}<g{m^2\over M_p^2}, \ee where
(\ref{condition}) is used. This indicates that the intensity of
the chirality oscillation is determined by the mass of scalar
field and its coupling $g$ to the gCS term. Generally, the rapid
oscillation requires $H<m\ll M_p$. Thus a larger $A_*$ implies a
larger $g$. In string theory, $g=\pi^2\sqrt{g_s/2}M_p^2/M_{s}^2$,
e.g., \cite{Alexander:2004wk}\cite{Alexander:2009tp}, where $M_s$ is
the string scale and $g_s$ is the string coupling, we may have
$g\gg 1$ for $M_s\ll M_p$. Thus the amplitude of oscillation in
CMB TB/EB spectrum may be straightly linked to the stringy
parameters.

\section{Discussion}

In summary, we found a novel phenomenon of the chirality of
gravity at inflation scale, i.e., the amplitudes of left- and
right-handed primordial GWs modes will convert into each other and
oscillate in their propagations. We illustrated it by applying the
gCS term coupling to a massive scalar field\footnote{We also
observed that when the gCS term is replaced by
$W\widetilde{W}$, in which $W$ is the Weyl tensor, the chirality
oscillation also appears. The result is similar. }. This chirality
oscillation will eventually develop a permanent difference of the
amplitudes of both modes, which leads to the nearly opposite
oscillating shapes in the left- and right-handed GWs spectrum.

The chirality oscillation of primordial GWs may bring obvious
wiggles in the CMB TB/EB-mode spectrum, which is the unique
fingerprint of chiral gravity.
Thus high-precision CMB B-mode polarization experiments could
offer us richer information on the UV-complete gravity theory than
expected, though the detecting is still a challenging issue
\cite{Gluscevic:2010vv}\cite{Zhao:2014yna}\cite{Gerbino:2016mqb}.

\textbf{Acknowledgments}

This work is supported by NSFC, No. 11222546, 11575188, and the
Strategic Priority Research Program of Chinese Academy of
Sciences, No. XDA04000000.

 \end{document}